# Stabilized free-space optical frequency transfer

D. R. Gozzard,[1,2,*] S. W. Schediwy,[1,2] B. Stone,[2] M. Messineo,[2] M. Tobar,[3]

[1]International Centre for Radio Astronomy Research, ICRAR M468, The University of Western Australia, 35 Stirling Hwy, Crawley, Western Australia, 6009
[2]Department of Physics, School of Physics, Mathematics & Computing, The University of Western Australia, 35 Stirling Hwy, Crawley, Western Australia, 6009
[3]ARC Centre of Excellence for Engineered Quantum Systems, Department of Physics, School of Physics, Mathematics & Computing, The University of Western Australia, 35 Stirling Hwy, Crawley, Western Australia, 6009

*david.gozzard@research.uwa.edu.au

**Abstract**

The transfer of high-precision optical frequency signals over free-space links, particularly between ground stations and satellites, will enable advances in fields ranging from coherent optical communications and satellite Doppler ranging to tests of General Relativity and fundamental physics. We present results for the actively stabilized coherent phase transfer of a 193 THz continuous wave optical frequency signal over horizontal free-space links 150 m and 600 m in length. Over the 600 m link we achieved a fractional frequency stability of $8.9 \times 10^{-18}$ at one second of integration time, improving to $1.3 \times 10^{-18}$ at an integration time of 64 s, suitable for transmission of optical atomic clock signals. The achievable transfer distance is limited by deep-fading of the transmitted signal due to atmospheric turbulence. We also estimate the expected additional degradation in stability performance for frequency transfer to Low Earth Orbit.

## I. INTRODUCTION

Free-space transfer of phase-coherent signals from optical atomic clocks, particularly between satellites and ground-stations, will enable significant improvements in coherent optical communications, timescale comparison over inter-continental distances, precision navigation, satellite Doppler ranging, geodesy, tests of General Relativity, and other fundamental physics experiments [1-11]. Free-space links can also offer advantages in cost and flexibility of deployment compared to optical fiber networks [12, 13]. However, refractive index fluctuations of air due to atmospheric turbulence induce phase perturbations on the transmitted optical signals. This degrades the coherence and therefore the usefulness of the signals as a scientific and industrial tool [4, 14-17]. Active phase stabilization systems can overcome this degradation and greatly improve the accuracy and precision of the transmitted signals.

Time and frequency transfer techniques previously developed for the distribution of high-precision signals over long distance optical fiber links [18-23] can also be applied to free-space transmission [16]. Two-way time and frequency transfer systems employing optical frequency combs at the transmit and receive ends have already demonstrated time and frequency transfer over horizontal links through up to 12 km of atmosphere [12-14] at levels of precision suitable for the synchronization and comparison of optical atomic clocks. A space-based optical atomic clock synchronized using ground to space links of such a level of stability would enable measurements of General Relativity, and other fundamental physics experiments, an order of magnitude more precise than will be achieved by the ACES mission [24-26]. However, the size and complexity of optical frequency combs restricts the types of satellites that these systems may be deployed on, and thus limiting the range of scientific and industrial applications of such satellites. In contrast, coherent, continuous wave, frequency transfer techniques require only relatively simple optics and detection hardware at the remote site and so offer advantages in terms of size, weight, and robustness that are critical to successful deployment of such a system on board a satellite. Optical Doppler ranging to satellites and measurements comparing the timescales of ground- and space-based optical clocks using continuous wave frequency transfer will also be able to achieve superior precision during the limited measurement window available as an orbiting spacecraft is within view compared with measurements made using two-way time and frequency transfer (pulsed) techniques [27, 28]. However, coherent frequency transfer with active phase stabilization has so far only been demonstrated over distances on the order of 100 m [29] and then only at microwave frequencies.

In this paper we present results of a simple system for the phase-stabilized transmission of a 193 THz (1550 nm) continuous wave optical signal over horizontal free-space links 150 m and 600 m in length, which are estimated, based on results in [30] and [31], to be equivalent to transmission distances of approximately 2.2 km and 9 km vertically through the atmosphere. We describe the design of the system's free-space terminals and show that deep-fading of the optical signal due to atmospheric turbulence limits the availability of coherent free-space optical transfer links using this system. Adaptive optics systems will be required to achieve reliable operation over horizontal links in excess of 1 km in length.

## II. EXPERIMENTAL SETUP

### A. Phase stabilized optical frequency transfer system

Phase stabilized frequency transfer techniques utilize a round-trip signal through the transmission link to measure and compensate for phase fluctuations that are caused by disturbances on the link. Because of this, stabilized frequency transfer systems require the transmission link to have a high degree of reciprocity in order to precisely measure the link-induced noise and maximize the suppression of phase fluctuations. The reciprocal behaviour of free-space links depends on details of the setup, such as the point-ahead angle, and a thorough analysis of this and other setups is covered in [16]. The experiments presented here transmit across a fixed length of turbulent atmosphere in which time-of-flight, atmospheric scintillation, and fading all exhibit reciprocal behavior [32-34], meaning that high-precision stabilized frequency transfer techniques are compatible with these free-space links.

The frequency transfer system used in these experiments is an evolution of the design pioneered by Ma and colleagues [23]. As shown in Figure 1, a 193 THz (1550 nm) optical signal from a laser passes through a 50/50 splitter where one half of the signal reflects from a Faraday mirror (FM), then back through the 50/50 splitter, and finally falls on a photodetector (PD). This forms the short arm of an imbalanced Michelson interferometer, thereby providing an optical reference signal for the phase stabilization system, with the long arm of the Michelson interferometer being the optical transmission link. The laser used in these experiments was an NKT Photonics Koheras X15 with a linewidth of < 100Hz, which provides a coherence length in excess of 950 km, sufficient to ensure that the linewidth of the laser does not affect the link stability performance even for transmissions to lower-altitude Low Earth Orbits. The other half of the laser signal passes through an acousto-optic modulator (AOM) where it is shifted by +50 MHz before it is passed to a collimator and beam expander, and is then launched into free-space. At the remote site, another beam expander and collimator are used to couple the arriving beam into a single mode fiber. The received signal is split with one half of the signal being directed to the end user or application. The other half of the signal passes through a second AOM which applies a +70 MHz frequency shift. This distinguishes the signal returning from the receiver from other unwanted reflections which can then be filtered out in the electronic domain. The signal then reflects from a FM, through the AOM again where it receives an additional +70 MHz, and back through the free-space link, returning to the transmitter unit. The signal returning to the transmitter unit passes through the +50 MHz servo AOM again and terminates on the PD where it generates a 240 MHz beat with the optical reference. The 240 MHz signal carries the phase fluctuations acquired by the transmitted optical signal. The 240 MHz signal is filtered and mixed with a 240 MHz local oscillator reference and the resulting DC signal is used to steer the frequency applied to the servo AOM to compensate for phase fluctuations on the link.

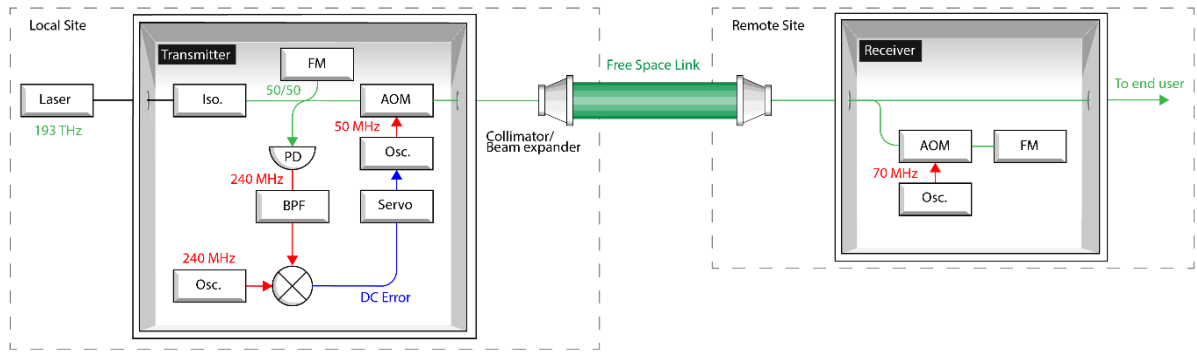

FIG. 1. (Color online) Free-space phase stabilized frequency transfer system. Red paths indicate electronic (radio frequency) connections, green paths indicate optical connections. Iso., isolator; PD, photo-detector; AOM, acousto-optic modulator; Osc., oscillator signal source; BPF, band-pass filter; FM, Faraday mirror.

### B. Folded beam link

The architecture of the stabilized frequency transfer system is intended to have the receiver unit installed in a remote location from the transmitter unit. However, in order to obtain a true measurement of the phase stability of the transmitted signal, the transmitter and receiver must be co-located so that the transmitted and received signals can be compared via an out-of-loop link. This was achieved by "folding" the free-space link using a corner cube reflector as shown in Figure 2. This means that only one optical terminal was used because the beam reflected by the corner cube reflector returned through the same transmitting optic. A 50/50 fiber splitter was used to couple the transmitter and receiver terminals to the single optic. This greatly simplified the setup of the test link but had the disadvantage that the transmitted signal lost 3 dB of power with every pass through the splitter, resulting in a total of 12 dB loss that would not be present on an operational free-space link.

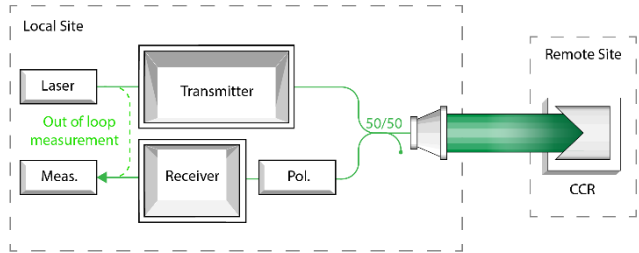

FIG. 2. (Color online) Folded architecture of experimental free-space link using a corner cube reflector. CCR, corner cube reflector; Pol., polarization controller.

The optical signal reaching the receiver unit was combined with an optical reference that was split off from the laser signal previously. The signals terminated on a measurement PD where they produced a 50 MHz signal that was used to measure the stability of the optical transmission. A polarization controller in the remote unit was used to align the polarization of the received light with the polarization of the optical reference from the laser. The 50 MHz signal was directed to an Agilent 53132A Λ-type high-precision frequency counter, which averaged the frequency over a 1 s gate time, and separately to a Microsemi 5125A phase noise test set with an acquisition rate of 2 MHz. The output of the counter was logged and used to compute an estimate of the fractional frequency stability of the transmitted signal for both stabilized and unstabilized transmission over the 150 m and 600 m horizontal links. The 5125A was used to measure the PSD of the signal phase noise.

**C. Optical terminal design**

In operation, two identical free-space optical terminals are required for the frequency transfer system, one at the transmitter site and one at the receiver site, but because of the folded beam path used in these tests, only one optical terminal was needed. Excluding the effects of atmospheric turbulence, the effective range of the free-space frequency transfer system will be limited by the optical power loss during transmission due to diffractive losses, extinction losses in the propagation medium, and loss due to coupling into the optical fiber. Because of this, the terminal was designed to be as low loss as reasonably practical.

The size of the terminal optics was determined by the desire to achieve a diffraction limited beam over the 600 m link, and by the Fried parameter which sets an upper limit on the size of the aperture for which the received light can be efficiently coupled into a single mode fiber [13, 35]. Data from [13] showed that an optic of 5 cm diameter was the largest that would achieve efficient coupling over 600 m in strong turbulence. Therefore, standard 2" (5 cm) optics provide a good solution for the design of the optical terminals and will also be useful for future trials over distances on the order of 10 km during moderate to light atmospheric turbulence conditions.

The final terminal design incorporated a fiber-to-free-space collimator that produced a $1/e^2$ intensity radius of 1.12 mm. This beam was directed into a 15:1 beam expander, resulting in a waist with a radius of 16.8 mm and with a divergence of 29 µrad. The beam expander had a nominal aperture of 50 mm, which was reduced to a clear aperture of 48 mm by the mounting hardware. The corner cube reflector (gold coated for efficient reflection at 1550 nm) had a clear aperture of 50.8 mm. These optics allow us to have diffraction limited divergence over the whole 600 m atmospheric link, producing a beam $1/e^2$ intensity radius of 24.4 mm returning to the beam expander aperture.

**D. Free-space links**

Two free-space links across the University of Western Australia (UWA) campus were used for these experiments (Figure 3). The stabilized frequency transfer system and free-space terminal were installed behind a window in an office on the 5th floor of the UWA Physics building (50 m above sea level, approximately 40 m above ground level). The corner cube reflector was mounted on a balcony in the Electrical Engineering building 75 m away, creating a 150 m test link, and behind a window at the UWA Oceans Institute building 300 m away, creating a 600 m test link. The link from Physics to Electrical Engineering had a 30° slant propagation path passing over a car park, while the link from physics to the Oceans Institute was nearly horizontal and passed over a car park and several campus buildings, approximately 25 m above the taller buildings. The link was achieved with both the Physics and Oceans Institute windows closed, which introduced additional attenuation to the transmitted signal.

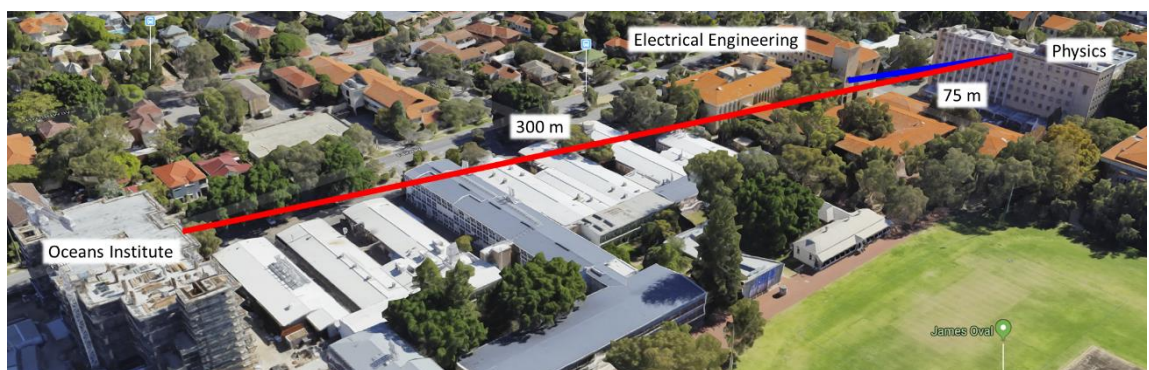

FIG. 3. (Color online) Beam path from Physics to Electrical Engineering (blue, 75 m) and to the Oceans Institute (red, 300 m). (Modified from Google Maps image.)

In both cases, the optical power launched into the collimator and beam expander was +6 dBm (approximately 4 mW). The coupling loss in and out of the fiber was 3 dB, and the corner-cube reflector produced an attenuation of approximately 2 dB due to the accumulation of dust on the mirrors. Including these losses, the 150 m link to Electrical Engineering had a loss of 13.5 dB (5.5 dB for the free-space path alone), while the 600 m link to the Oceans Institute had a loss of 15.5 dB (7.5 dB for the free-space path alone). The expected losses due to scattering and absorption of the 1550 nm light in the air column are approximately 0.5 dB and 2.5 dB for the 150 m and 600 m links respectively. (This figure assumes attenuation due to the moderate haze typical of this location and time of year and includes a slight clipping of the returning beam for the 600 m link [36, 37].) We attribute the much higher than expected free-space path power loss to the fact that the beam passed through windows at a significant angle. For the link from Physics to Electrical Engineering, the beam encountered one window at a shallow angle of approximately 30°, while the link to the Oceans Institute passed through two windows at a steeper angle of approximately 75°. Significant reflection and attenuation of the beam occurred with each pass. Cleaning the Physics window reduced losses by 0.5 dB. We expect some additional parasitic losses in the fiber connectors between the free-space terminal and the power meter. A further 3 dB of loss was acquired due to the optical splitter used to couple the transmitter and receiver to the single collimator and beam expander. The optical power entering the receiver was then −10.5 dBm (0.09 mW) for the 150 m link and −12.5 dBm (0.06 mW) for the 600 m link. The loss within the receiver unit is 12 dB. After traveling through the link again, and including two passes of the optical splitter, the optical power returning to the transmitter side was −42 dBm (0.063 µW) for the 150 m link and −46 dBm (0.025 µW) for the 600 m link.

It should be noted that the fiber splitter used to connect the transmitter and receiver units to the free-space optic produced a reflection at around −35 dBm that was indistinguishable to the stabilization system from the reflection from corner-cube. The power of this unwanted reflection ultimately determines the limit for the link power loss for this experiment when the system is used in this folded beam path setup. However, this reflection would not be present in normal operation and would not limit the range or performance of an operational stabilization system.

## III. RESULTS

The fractional frequency stability results computed from the frequency counter data are shown in Figure 4. The blue curves with diamond markers represent the stability of the 150 m link, while the red curves with circle markers represent the stability of the 600 m link. The dashed lines with open markers are measurements of the unstabilized frequency transfer and the solid lines with filled markers are the measurements of the stabilized transfer.

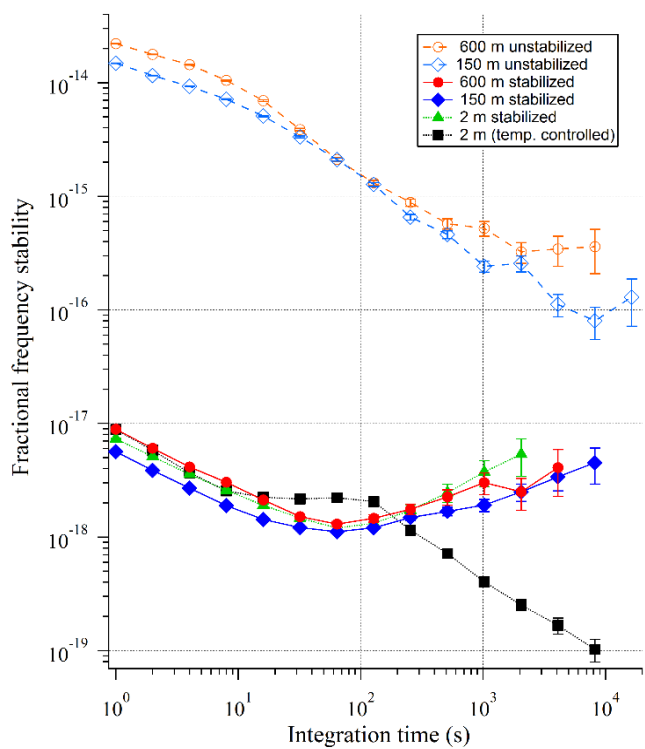

FIG. 4. (Color online) Fractional frequency stability of the 193 THz optical signal transmitted over 150 m (blue diamonds) and 600 m (red circles). Filled markers with solid lines indicate stabilized transmission. Open markers with dashed lines indicate unstabilized transmission. Black squares — transmission over 2 m in a temperature-controlled laboratory. Green triangles — transmission over 2 m in the Physics building office.

The length of the data acquisition for the 600 m link was limited by the magnitude of the atmospheric turbulence. During periods of high turbulence, usually as the ground and buildings heated up in the morning sunlight, scintillation and beam wander of the laser in the atmosphere caused instances of deep fading of the optical power, causing brief losses of lock and breaks in the frequency counter data. The measurements presented in Figure 4 are for the longest break-free period of the data acquisition, about 10 hours, during a period when the atmosphere was relatively stable. Each set of measurements, 150 m stabilized, 150 m unstabilized, 600 m stabilized, and 600 m unstabilized, was continued for five days to observe how the performance of the system varied with changing conditions. The results in Figure 4 are typical of the general performance levels of the system during the testing period. The measured stabilized signal stability varied by a factor of about two depending on weather conditions.

Over the 150 m link the stability of the unstabilized transmission was $1.5 \times 10^{-14}$ at an integration time, $\tau$, of one second and integrated down over time. With the stabilization system engaged, the stability was improved by more than three orders of magnitude, achieving $5.7 \times 10^{-18}$ at one second. The stability improves with a $\tau^{-1/2}$ trend to $1.1 \times 10^{-18}$ at an integration time of 64 s before becoming degraded at longer integration times, where the trace exhibits a $\tau^{1/2}$ trend.

The unstabilized 600 m transmission had a stability of $2.2 \times 10^{-14}$ at one second and also continued to integrate down over time. The stabilization system improved the stability by more than three orders of magnitude, achieving $8.9 \times 10^{-18}$ at one second. Like the 150 m link, the 600 m link exhibits a $\tau^{-1/2}$ trend up to $\tau = 64$ s where it achieved a best stability of $1.3 \times 10^{-18}$ before the stability degraded at longer integration times, displaying a $\tau^{1/2}$ trend.

Also included in the plot in Figure 4 are results for the stability of the system over short 2 m free-space test links. The black trace with square markers is for a test performed over 2 m in an airconditioned metrology laboratory, while the green trace with triangular markers is for a test over 2 m in the unregulated temperature conditions of the un-airconditioned Physics 5th floor office. The results of the 2 m laboratory tests (black squares) show a general $\tau^{-1/2}$ trend up to long integration times, while the results for the 5th floor office (green triangles) exhibit the same change at $\tau = 64$ s as the results from the 150 m and 600 m links.

Figure 5 shows plots of the PSDs for the 150 m and 600 m links, as well as for a 'zero-length' test achieved by mounting a corner-cube reflector directly at the output of the free-space terminal. For both the 150 m and 600 m links, the stabilized frequency transfer improves the phase noise by 60 dB at an offset frequency of 0.01 Hz. This tapers to an improvement of 50 dB at 30 Hz for the 150 m link, and 40 dB at 30 Hz for the 600 m link.

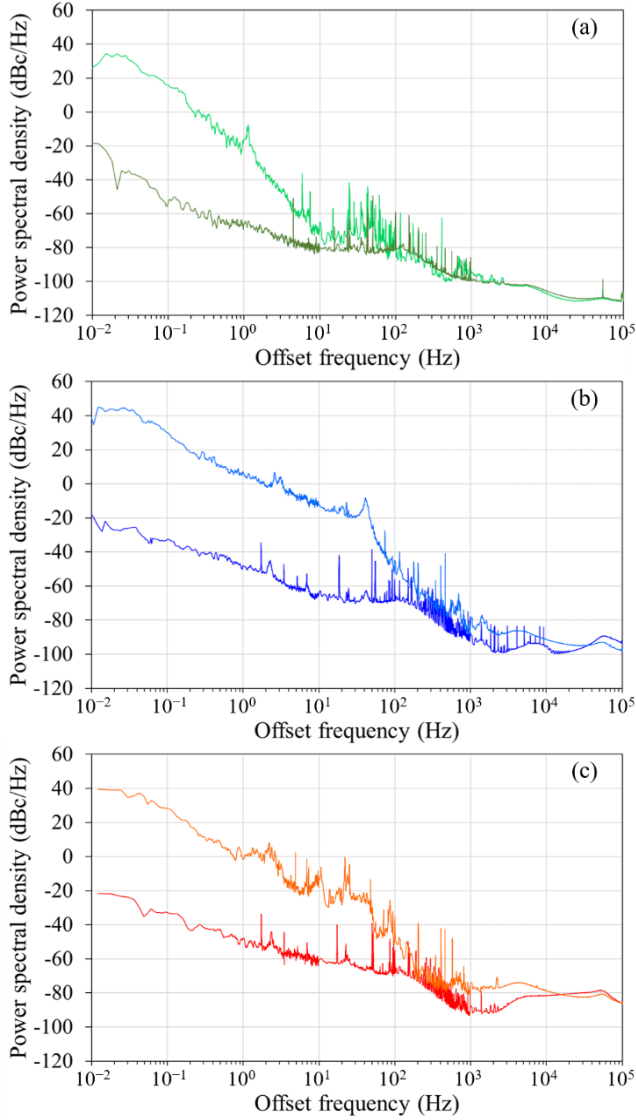

FIG. 5. (Color online) PSD of 193 THz optical signal transmitted over (a) 0 m, (b) 150 m, and (c) 600 m. Dark green — 0 m stabilized, light green — 0 m unstabilized, dark blue — 150 m stabilized, light blue — 150 m unstabilized, red — 600 m stabilized, orange — 600 m unstabilized.

## IV. DISCUSSION

The fractional frequency results (Figure 4) show that the stabilized transfer system suppresses the free-space link noise by more than three orders of magnitude. Over these atmospheric links, we achieved levels of fractional frequency stability, at integration times less than 64 s, comparable to the best achieved over optical fiber links of comparable length [5, 21], and exceeded the stability achieved by [12] and [38] using two-way time and frequency transfer of an optical comb over 2 km atmospheric links by an order of magnitude.

The Agilent 53132A frequency counter used to perform these measurements exhibits an approximately 0.1 s dead-time between each 1 s frequency averaging period which, when successive measurements are juxtaposed in order to average over a longer integration time, biases the expected $\tau^{-1}$ slope of the fractional frequency data to $\tau^{-1/2}$ [39-41]. This means that the $\tau^{-1/2}$ trend of the data up to $\tau = 64$ s indicates that the dominant noise process at these shorter integration times is consistent with white phase noise. That the dominant noise process is white phase noise was confirmed using the data from the Microsemi test set. Beyond τ = 64 s, the fractional frequency values increase due to temperature changes, mainly diurnal temperature variations.

The results of the short 2 m link tests in the office and the laboratory show that this degradation of the out-of-loop measurement of the fractional frequency stability after 64 s is due to temperature changes within the room that the tests were performed in. The room was an ordinary office with poor temperature stability and no airconditioning, rather than an airconditioned laboratory. This meant that the out-of-loop measurement system had poor thermal isolation resulting in a degradation of the measured frequency stability at timescales above 64 s. Better thermal isolation of the out-of-loop components will improve the measured performance of the system at longer integration times. From the results of the 2 m tests in the airconditioned laboratory we expect the actual stability of the system over the 150 m and 600 m links to reach the $1\times10^{-19}$ level by $10^4$ seconds of integration. As this is a free-space frequency transfer system, the optical terminals will, by necessity, be relatively exposed to the elements in operation. However, as the entire link (long arm of the Michelson interferometer) is stabilized, the stabilization system can be placed at the nearest convenient room with good temperature stability and a fiber patch link run to the optical terminal.

The PSD plots for the unstabilized links, Figure 5, show that the atmosphere is the dominant source of phase noise at offset frequencies from 0.1 Hz to several hundred Hz. Below 0.1 Hz, a large fraction of the phase noise is contributed by mechanical disturbances, including low-frequency vibrations and temperature fluctuations, to the free-space frequency transfer system. This suggests that, in these tests where the equipment was being used in a relatively noisey room with poor temperature stability, the low-frequency performance of the stabilization system is close to being limited by these mechanical disturbances. The stabilized frequency transfer achieves up to 60 dB improvement in phase noise at low offset frequencies and continues to provide effective suppression at offset frequencies up to several hundred Hz over both the 150 m and 600 m links. This is consistent with the atmospheric turbulence model in [35] that predicts that the angle of arrival jitter spectrum cuts off sharply at frequencies above $10V/2\pi D$ (where V is the wind speed in m/s and D is the aperture diameter in meters). The phase fluctuations at frequencies higher than 1 kHz seen in the plots, frequencies where atmospheric fluctuations are unlikely to have a significant contribution, are most likely due to acoustic frequency mechanical disturbances to the stabilized transfer system and to the fiber patch that delivers the optical signal to the free-space terminal. The increase in phase noise above 1 kHz from zero-length, to 150 m, to 600 m is due to the slightly different setups for each measurement. For example, to achieve the 600 m link, the free-space terminal needed to be mounted to a different wall, and have a longer fiber run to it from the stabilization system, than for the 150 m link, while the zero-length test was achieved with a corner-cube reflector mounted rigidly to the free-space terminal, avoiding any differential motion between the terminal and reflector. The demonstrated suppression of thermal and seismic noise, as well as atmospheric noise, by the stabilized frequency transfer system shows that the system is capable of handling real-world applications where the equipment may not be installed in ideally quiet environments, and the transmitter and receiver units will be subject to differential motion

The data for the unstabilized links in panels (b) and (c) of Figure 5 exhibit slopes close to $f^{-8/3}$ between offset frequencies from 0.1 Hz to 50 Hz, consistent with the noise spectrum that would be expected from standard Kolmogorov turbulence theory [4, 42]. However, the traces for the unstabilized links in Figure 4 do not exhibit the $\tau^{-1/6}$ slope expected from the theory. Instead, the traces exhibit slopes close to $\tau^{-0.36}$ between $\tau = 1$ s and $\tau = 8$ s where they start to exhibit slopes of $\tau^{-1}$ until long-period temperature fluctuations begin to dominate at $\tau = 512$ s. This discrepancy between the measured values and that expected from the theory is due to the dominance of noise due to mechanical perturbations as seen in Figure 5.

The atmosphere was found to be most stable between the times of 5pm and 5am. PSD measurements were obtained during the early evening, with fractional frequency stability measurements being left to accumulate data over night until increased atmospheric turbulence in the late morning caused signal fading and a resulting loss of lock and cycle slips. Temperature differentials during the day would create much greater turbulence, to the extent that on fine mornings with little wind, the turbulence was so great that deep-fading of the link signal would occur regularly, causing the system to lose lock every few seconds on the 600 m link. On a typical day, the 600 m link would experience less than one cycle slip per hour between 5pm and 5am. The rate of cycle slips would increase to an average of about one cycle slip per minute at midday. Days with extremely bad turbulence could produce a cycle slip every five seconds on average, which could increase still further to one cycle slip per second if it was raining. It is because of these cycle slips that the Agilent 53132A was the device primarily used to obtain the fractional frequency stability data, despite exhibiting measurement dead time, because it would continue recording after a cycle slip, whereas the Microsemi 5125A would abort the measurement if a cycle slip or significant signal power reduction occurred. This meant that the Agilent device could be used to record the system performance over longer periods of time.

Overall, the stability levels achieved by the system shown in both the PSD and fractional frequency stability plots are suitable for applications where comparison between optical atomic clocks is required. However, atmospheric scintillation causing signal fading became so pronounced on the 600 m link that such a link becomes un-useable in adverse turbulence conditions. Because of this, 600 m seems to be near the longest useful horizontal link achievable without the inclusion of adaptive optics systems. Indeed, [14] reported in excess of 100 signal interruptions per second over their 12 km free-space optical link, which would severely limit the availability of coherent frequency transfer links.

Both [4] and [12] report that the integrated turbulence across their respective 2 km and 5 km links is similar to the integrated turbulence from ground to a satellite. While our longest link was only 600 m, this means we were sampling atmospheric noise with a PSD within an order of magnitude of that expected for a vertical path through the whole atmosphere. The fractional frequency and PSD results reported in Figures 4 and 5 respectively show that the performance of the stabilized transfer system is not yet distance limited, so similar performance over longer horizontal and vertical links should be possible with the addition of an adaptive optics system. Extending the system to a 12 km link, equivalent to those achieved by [13] and [14], we would expect the free-space signal to be exposed to greater phase noise, however, the servo loop bandwidth, which is set by the signal's round-trip time, would still be greater than 10 kHz. From Figure 5 we see that this reduction in loop bandwidth should not result in a significant increase in the amount of phase noise that is outside the servo bandwidth, and which would not be suppressed by the servo. Therefore, we expect the servo loop to continue to effectively suppress the in-loop phase noise and thus the fractional frequency stability performance of an approximately 12 km horizontal link to be similar to the current result.

The optical power budget with the current optical terminals is insufficient to allow stabilized transfer to Low Earth Orbit. Higher initial optical power will extend the range of the present system, and adaptive optics will also enable the use of larger transmitting and receiving optics. Even though the integrated noise on a 12 km horizontal link should be equal to or less than that of a vertical link to Low Earth Orbit, the reduction in feedback bandwidth due to the light round-trip time should begin to impact the achievable stability over the distances involved in transmitting to Low Earth Orbit and beyond.

For an approximately 1000 km link to Low Earth Orbit, the servo bandwidth would be reduced to 150 Hz. As shown in Figure 5, the dominant phase noise induced by atmospheric fluctuations occurs at frequencies lower than approximately 100 Hz and so there would only be a moderate increase in out-of-loop phase noise. This can be compared to the 1100 km stabilized optical frequency transfer via fiber optic link implemented in [43] in which they achieved a fractional frequency stability of $4\times10^{-16}$ at one second with a servo bandwidth of 100 Hz (due to the lower speed of light in fiber). However, because the fiber link picks up phase noise along its entire length, unlike a free-space link to orbit in which the majority of the phase noise would be acquired in the first few kilometres, the unstabilized phase noise in [43] is much greater than for a vertical free-space link. (See [43] Figure 4 (b).) We therefore expect the fractional frequency performance of stabilized frequency transfer to Low Earth Orbit to remain below the $4\times10^{-16}$ at one second value reported by [43].

The coherence length of the laser used in these tests was 950 km, but real applications of this technology will use either a cavity stabilized laser (linewidth < 1 Hz) or a laser locked to an optical atomic clock, which will enable phase-stable optical frequency transfer throughout Low-Earth Orbit and beyond.

## V. CONCLUSION

The stabilized optical frequency transfer system reported in this paper effectively suppressed the phase perturbations induced on the signal by atmospheric turbulence and refractive index changes, achieving levels of stability over these short distances comparable to the best stabilized fiber links, and exceeded the fractional frequency stability achieved by [12] and [38] over 2 km free-space links by an order of magnitude. By either improving the thermal isolation of the stabilized frequency transfer system or locating the system in a temperature stable room away from the optical terminal, we expect the stability performance to continue to integrate down, as in our laboratory-based measurement, over a period of at least several minutes typical of the transit times of satellites in Low Earth Orbit. Work is underway to improve the thermal shielding of the frequency transfer system.

Taking into account the reduction in servo loop bandwidth and comparing the expected (lower) phase noise of a vertical free-space link with that of an optical fiber link of comparable distance, we expect the fractional frequency performance to Low Earth Orbit to be better than the $4\times10^{-16}$ at one second achieved by [43], and to integrate down (at $\tau^{-1/2}$ if using an Agilent 53132A) over a period of several minutes. This makes the system suitable for use in the transmission of optical atomic clocks signals and other high precision coherent signals. The use of only small size and low complexity components at the receiver, makes this system ideally suited for deployment on future spacecraft for ground-to-space optical timescale comparison.

Currently, the practical horizontal transmission distance is limited to about 600 m by deep fading of the optical signal caused by scintillation and jitter in beam angle-of-arrival. To achieve effective and useful transmission of coherent signals over longer distances, adaptive optics systems will be needed to correct the wavefront aberrations that cause this deep fading.

Adaptive optics systems have previously been shown to be compatible with stabilized optical two-way time and frequency transfer systems [16, 44]. We are currently exploring combining the stabilized frequency transfer system with adaptive optics. Preliminary results over a 22 m link in the laboratory indicate that adaptive optics systems are compatible with this technique and have shown to lead to further improvements in fractional frequency stability and noise PSD.

With the inclusion of an adaptive optics system we expect the transfer distance of the system to be limited by the optical power loss due to beam divergence. Under this scenario, the range of the system can be extended by the inclusion of erbium-doped fiber amplifiers (EDFAs) at the terminal sites to compensate for the optical power loss of the long links. EDFA's preserve the reciprocity of the link and have been used effectively in fiber-optic links for the stabilized transmission of optical clock signals [5, 21]. Based on results in [30] and [31], with the inclusion of an adaptive optics system we expect the current optical terminal design to be able to achieve a stabilized transfer distance on an operational horizontal link of approximately 40 km in clear air, and up to 100 km vertically through the atmosphere. A more powerful laser at the local site or larger optics can be used to improve the link power budget and achievable range. For example, increasing the diameter of the optics by a factor of two would increase the effective transfer range (with suitable adaptive optics) to approximately 400 km.

## ACKNOWLEDGEMENTS

The authors wish to thank Dr. Francis Bennet and Dr. Lyle Roberts for their efforts in enabling the preliminary adaptive optics tests, and Madeleine Sheard for her assistance with this work. This work was supported by the Australian Research Council's Linkage Infrastructure, Equipment and Facilities (LE160100045) funding scheme and the ARC Centre of Excellence in Engineered Quantum Systems (CE170100009).